\documentclass[intlimits,twoside,a4paper]{article}

\usepackage{amsmath,amssymb}
\usepackage{graphicx}
\usepackage{wrapfig}

\usepackage[T2A]{fontenc}
\usepackage[cp1251]{inputenc}

\usepackage[eqsecnum]{cmpj2}

\issue{2012}{15}{4}{43006}

\doinumber{10.5488/CMP.15.43006}

\newcommand{\non}{\nonumber \\}
\newcommand{\be}{\begin{equation}}
\newcommand{\ee}{\end{equation}}
\newcommand{\bea}{\begin{eqnarray}}
\newcommand{\eea}{\end{eqnarray}}
\newcommand{\lp}{\left (}
\newcommand{\rp}{\right )}
\newcommand{\lbr}{\left [}
\newcommand{\rbr}{\right ]}
\newcommand{\lb}{\left \{}
\newcommand{\rb}{\right \}}
\newcommand{\ld}{\left .}
\newcommand{\rdo}{\right .}
\newcommand{\ve}[1]{{\bf #1}}
\newcommand{\vk}{\ve{k}}
\newcommand{\rhok}{\rho_{\ve{k}}}
\newcommand{\rhomk}{\rho_{-\ve{k}}}

\newcommand{\vl}{{\ve{l}}}
\newcommand{\cH}{{\cal H}}
\newcommand{\cW}{{\cal W}}
\newcommand{\cF}{{\cal F}}
\newcommand{\cB}{{\cal B}}
\newcommand{\cA}{{\cal A}}
\newcommand{\cK}{{\cal K}}
\newcommand{\cC}{{\cal C}}
\newcommand{\cL}{{\cal L}}
\newcommand{\cD}{{\cal D}}

\title[Critical behaviour of a 3D Ising-like system]%
{Critical behaviour of a 3D Ising-like system \\ in the $\rho^6$
model approximation:  Role of the correction \\ for the potential averaging}
\author[I.V. Pylyuk, M.V. Ulyak]{I.V. Pylyuk\refaddr{label1}\thanks{E-mail:
        piv@icmp.lviv.ua}\,, M.V. Ulyak\refaddr{label2}}
\addresses{
\addr{label1} Institute for Condensed Matter Physics of the National
Academy of Sciences of Ukraine, \\ 1 Svientsitskii St., 79011 Lviv, Ukraine
\addr{label2} Chervonograd State College of Mining Technologies
and Economics,  17 Stus St., 80100 Chervonograd, Ukraine
}

\authorcopyright{I.V. Pylyuk, M.V. Ulyak, 2012}

\date{Received July 3, 2012, in final form September 12, 2012}

\begin{document}

\maketitle

\begin{abstract}
The critical behaviour of systems belonging to the three-dimensional
Ising universality class is studied theoretically using the collective
variables (CV) method. The partition function of a one-component spin
system is calculated by the integration over the layers of the CV phase
space in the approximation of the non-Gaussian sextic distribution of
order-parameter fluctuations (the $\rho^6$ model). A specific feature of
the proposed calculation consists in making allowance for the dependence
of the Fourier transform of the interaction potential on the wave
vector. The inclusion of the correction for the potential averaging leads to
a nonzero critical exponent of the correlation function $\eta$ and
the renormalization of the values of other critical exponents.
The contributions from this correction to the recurrence relations for
the $\rho^6$ model, fixed-point coordinates and elements of the
renormalization-group linear transformation matrix are singled out.
The expression for a small critical exponent $\eta$ is obtained in
a higher non-Gaussian approximation.
\pacs 05.50.+q, 64.60.F-, 75.10.Hk
\end{abstract}

\section{Method}

We shall use the approach of collective variables
(CV)~\cite{ymo287,ykpmo101}, which allows us to calculate the expression
for the partition function of a system and to obtain complete
expressions for thermodynamic functions near the phase-transition
temperature $T_\mathrm{c}$ in addition to universal quantities (i.e., cri\-ti\-cal
exponents). The CV approach is non-perturbative and similar to
the Wilson non-perturbative renormalization-group (RG) approach
(integration on fast modes and construction of an effective theory for
slow modes)~\cite{btw102,tw194,bb101}.

The term collective variables is a common name for a special class of
variables that are specific for each individual physical system.
The CV set contains variables associated with order pa\-ra\-me\-ters.
Consequently, the phase space of CV is most natural in describing
a phase transition. For magnetic systems, the CV $\rhok$ are
the variables associated with the modes of spin-moment
density oscillations, while the order parameter is related
to the variable $\rho_0$, in which the subscript ``0'' corresponds to
the peak of the Fourier transform of the interaction potential.
The methods available at present, make it possible to calculate universal
quantities to a quite high degree of accuracy (see, for example,
\cite{pv102}). The advantage of the CV method lies in the possibility to
obtain and analyse thermodynamic characteristics as functions of
microscopic parameters of the original system.
The use of the non-Gaussian basis distributions of fluctuations in
calculating the partition function of a system does not bring about a
problem of summing various classes of divergent
(with respect to the Gaussian distribution) diagrams at
the critical point. A consideration of the increasing number of terms
in the exponent of the non-Gaussian distribution is an alternative
to the use of a higher-order perturbation theory based
on the Gaussian distribution.

The integration of partition function begins with the
variables $\rho_{\bf k}$ having a large value of the wave vector $k$
(of the order of the Brillouin half-zone boundary) and terminates
at $\rho_{\bf k}$ with $k \rightarrow 0$. For this purpose, we divide
the phase space of the CV $\rho_{\bf k}$ into layers with the division
parameter $s$. In each $n$th layer (corresponding to the region of
wave vectors $B_{n+1} < k \leqslant B_n$, $B_{n+1} =B_n/s$, $s>1$), the Fourier
transform of the interaction potential is replaced by its average value.
This simplified procedure leads to a zero value of the critical
exponent $\eta$ characterizing the behaviour of the pair-correlation
function at the critical temperature $T_\mathrm{c}$.

\section{The setup}

The object of investigation is a three-dimensional (3D) Ising-like
system. The Ising model, despite its simplicity, has, on the one hand,
a wide scope of realistic applications, and, on the other hand, it can
be considered as a model, which serves as a standard in studying other
models possessing a much more complicated construction.

In our previous calculations (for example,~\cite{pk191,ykp202,
kpp606,pk410}), we assumed that the correction for the potential
averaging is zero. As a result, we lost some information in the process
of calculating the partition function of the system. In particular,
the critical exponent $\eta$ was equal to zero.

\begin{figure}[!b]
\centerline{
\includegraphics[width=0.5\textwidth]{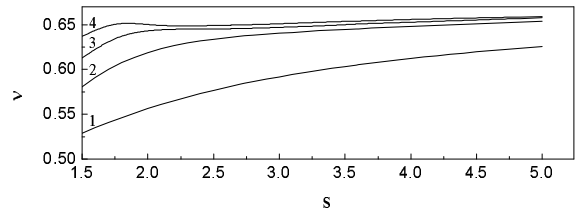}
}
\caption{Evolution of the critical exponent of the correlation
length $\nu$ with an increasing parameter of division of the CV phase space
into layers $s$. Curves~1, 2, 3 and 4 correspond to the
$\rho^4$, $\rho^6$, $\rho^8$ and $\rho^{10}$ models, respectively.}
\label{cmpf1}
\end{figure}

In~\cite{ykp312}, the correction for the averaging of the Fourier
transform of the potential is taken into account in the simplest
non-Gaussian approximation (the $\rho^4$ model based on the quartic
fluctuation distribution). The inclusion of this correction gives rise to
a nonzero value of the critical exponent $\eta$. Recurrence
relations (RR) between the coefficients of the effective distributions
take another form as compared with the case when $\eta=0$. A fixed
point is shifted. Critical exponents of the correlation length $\nu$,
susceptibility $\gamma$ and specific heat $\alpha$
are renormalized. Critical amplitudes are also modified. As is seen
from table~\ref{cmpt1}, the inclusion of a nonzero exponent $\eta$
\begin{table}[htbp]
\caption{Estimates of the critical exponents for the $\rho^4$ model and
the RG parameter $s=4$ in the case of the correction for the potential
averaging not taken into account ($\Delta\tilde\Phi(k)=0$) and in
the case of the correction for the potential averaging taken
into account ($\Delta\tilde\Phi(k)\neq 0$).}
\label{cmpt1}
\vspace{2ex}
\begin{center}
\begin{tabular}{|ccccc|}
\hline
\multicolumn{1}{|c}{Condition}
& \multicolumn{1}{c}{$\eta$}
& \multicolumn{1}{c}{$\nu$}
& \multicolumn{1}{c}{$\gamma$}
& \multicolumn{1}{c|}{$\alpha$} \\
\hline\hline
$\Delta\tilde\Phi(k)=0$ & 0 & 0.612 & 1.225 & 0.163 \\
$\Delta\tilde\Phi(k)\neq 0$ & 0.024 & 0.577 & 1.141 & 0.268 \\
\hline
\end{tabular}
\end{center}
\end{table}
within the CV method reduces the critical
exponent $\nu$ (like in the non-perturbative RG
approach~\cite{cdm203}). In order to obtain better quantitative
estimates of $\nu$ and other critical exponents, it is necessary
to use the distributions of fluctuations more complicated than
the quartic distribution. In the case of $\eta=0$, the critical
exponent of the correlation length for these distributions takes on
larger values than $\nu$ for the $\rho^4$ model (figure~\ref{cmpf1})
\cite{ykpmo101,ykp202}. The results of calculations and their comparison
with the other authors' data show that the sextic distribution (the $\rho^6$
model) provides a more adequate quantitative description of the critical
behaviour of a 3D Ising ferromagnet than the quartic
distribution~\cite{ykpmo101,p699}. The sextic distribution
for the modes of spin-moment density oscillations is presented as an
exponential function of the CV whose argument includes the sixth power
of the variable in addition to the second and the fourth powers.

In the present publication, our aim is to investigate the effect of
the above-mentioned correction for the potential averaging on the
critical properties of a 3D Ising-like system and to elaborate a
technique for calculating the small critical exponent $\eta$
in the $\rho^6$ model approximation. The analytic results, obtained in
this higher non-Gaussian approximation, provide the basis for accurate
analysis of the behaviour of the system near $T_\mathrm{c}$ with allowance for
the exponent $\eta$.

The developed approach permits to perform the calculations for a
one-component spin system in real 3D space on the microscopic level
without any adjustable parameters. The calculation technique for $\eta$
is similar to that proposed in the case of the quartic
distribution~\cite{ykp312}. New special functions appearing
in the construction of the phase-transition theory using the sextic
distribution were considered in~\cite{ykpmo101,p196}. In the case of
the $\rho^6$ model, we exploit the special functions with
two arguments more complicated than the parabolic cylinder functions with one argument
for the $\rho^4$ model.

\section{Basic relations}

We consider a 3D Ising-like system on a simple
cubic lattice with $N$ sites and period $c$. The Hamiltonian of such
a system has the form
\be
H = -\frac{1}{2} \sum_{\ve{i},\ve{j}} \Phi\left(r_{\ve{i}\ve{j}}\right)
s_\ve{i} s_\ve{j} - h \sum_\ve{i} s_\ve{i}\,,
\label{cmp1d1}
\ee
where $s_\ve{i}$ is the operator of the $z$-component of spin at
the $\ve{i}$th site, having two eigenvalues $+1$ and $-1$. The interaction
potential is an exponentially decreasing function
\be
\Phi\left(r_{\ve{i}\ve{j}}\right) = A \exp \left(-r_{\ve{i}\ve{j}}/b\right).
\label{cmp1d2}
\ee
Here, $A$ is a constant, $r_{\ve{i}\ve{j}}$ is the interparticle distance,
and $b$ is the radius of effective interaction.
In the representation of the CV $\rhok$, the partition function of the
system in the absence of an external field $\mbox{\boldmath $\cH$}$
($h=\mu_\mathrm{B}\cH=0$, $\mu_\mathrm{B}$ is the Bohr magneton) can be written
in the form
\bea
Z& = &2^N 2^{(N_1-1)/2} Q_0 [Q(P)]^{N_1}
\int \exp \lb - \frac{1}{2} \sum_{k\leqslant B_1}
\lbr d'(k) - d'\left(B_1,B'\right) \rbr \rhok \rhomk \rb \nonumber \\
& & \times \left(1+\hat\Delta_g+\cdots\right)
\exp \lbr - \frac{1}{2} R_2 \sum_{k\leqslant B_1} \rhok\rhomk \rdo \non
& & \ld - \sum\limits_{l=2}^3 \frac{R_{2l}}
{(2l)!N_1^{l-1}} \sum_{k_1,\ldots,k_{2l}\leqslant B_1}
\rho_{\vk_1}\cdots\rho_{\vk_{2l}}
\delta_{\vk_1+\cdots+\vk_{2l}} \rbr
(\rd\rho)^{N_1}\, ,
\label{cmp2d3}
\eea
where $B_1=B'/s$, $N_1=N's^{-3}$, $B'=\left(b\sqrt 2\right)^{-1}$, $N'=Ns_0^{-3}$,
$s_0=B/B'$, $B=\pi/c$ is the boundary of Brillouin half-zone, and
$\delta_{\vk_1+\cdots+\vk_4}$ is the Kronecker symbol. For the
coefficient $d'(k)$, we have
\be
d'(k) = a'_2 - \beta \tilde \Phi(k).
\label{cmp1d4}
\ee
Here, $\beta=1/(kT)$ is the inverse temperature. For
the Fourier transform of the interaction potential, we use the following
approximation~\cite{ykp202,rev9789}:
\be
\tilde\Phi(k) = \left\{
\begin{array}{ll}
\tilde\Phi(0) \left(1-2b^2k^2\right), & k\leqslant B', \\
0, & B'<k\leqslant B.
\end{array}
\rdo
\label{cmp1d5}
\ee
The quantities $Q_0$, $Q(P)$ and $R_{2l}$ are ultimate functions of the
initial coefficients $a'_{2l}\,(l=0,1,2,3)$~\cite{ykpmo101,p398}.
The coefficients $a'_{2l}$ are determined by special functions
of two arguments and are dependent on the ratio of the effective
interaction radius $b$ to the lattice constant $c$, i.e., on the
microscopic parameters of the system (see, for example,~\cite{kpd297}).

The correction, which is introduced by the operator
$\hat\Delta_g$, is considered in the linear approximation in
\be
\Delta\tilde\Phi(k) = q - 2 b^2 \beta\tilde\Phi(0) k^2.
\label{cmp2d2}
\ee
The quantity $\Delta\tilde\Phi(k)$ corresponds to the deviation
$\beta\tilde\Phi(k)$ from the average value $\beta\tilde\Phi \lp B_1,B'\rp $.
Here, $q=\bar q\beta\tilde\Phi(0)$, $\bar q$ defines the geometric mean
value of $k^2$ on the interval $(1/s,1]$. In the above-mentioned
approximation, we arrive at the expression
\bea
\hat\Delta_g^{(1)}& = &\frac{1}{2} \sum_{k_1,\ldots,k_6\leqslant B_1}
\left[ \frac{4C(h,\alpha)}{a'_4}\right]^2 \frac{\partial^6}
{\partial \rho_{\vk_1}\cdots\partial \rho_{\vk_6}} (N')^{-4}
\sum_{B_1<k\leqslant B'} \Delta g(k) \nonumber \\
& & \times  \sum_{\vl_1,\vl_2} \exp
\left[ -\ri\left(\vk_1+\vk_2+\vk_3+\vk\right)\vl_1-
\ri\left(\vk_4+\vk_5+\vk_6-\vk\right)\vl_2 \right],
\label{cmp2d10}
\eea
which defines the operator $\hat\Delta_g$ accurate to
within the term proportional to
$\frac{\partial^6}{\partial \rho_{\vk_1}\cdots\partial \rho_{\vk_6}}$
(the terms proportional to the higher orders of operators
$\partial/\partial \rhok$ are not taken into account).
The summation over the sites $\vl_1$, $\vl_2$ in (\ref{cmp2d10}) is
carried out for the lattice with period $c'=\pi b \sqrt 2$. The role of
$\Delta g(k)$ is played by the quantity
\be
\Delta g(k) = \frac{\Delta\tilde\Phi(k)}{1-S_2(2\pi)^{-2}\Delta\tilde\Phi(k)}\,,
\label{cmp2d8}
\ee
where $S_2=(2\pi)^2\left(24/a'_4\right)^{1/2}F_2(h,\alpha)$.
The forms of the functions $C(h,\alpha)$ and $F_2(h,\alpha)$ as well as
of their arguments $h$ and $\alpha$ are presented in the next section.
We assume that $\hat\Delta_g^{(1)}$ operates only
on \linebreak $\exp\left(-\frac{1}{2}R_2\sum_{k\leqslant B_1}\rhok\rhomk\right)$ in
(\ref{cmp2d3}). This assumption is associated with small contributions
from $R_4$ and $R_6$ in comparison with the contribution from $R_2$
(in particular, $R_{4}/\lp 6R_{2}^{2}\rp \sim 10^{-4}$~\cite{ykp312}).

\section{Partition function of the system with allowance for
the correction for the potential averaging}

The integration over the zeroth, first, second, \ldots, $n$th layers
of the CV phase space leads to the representation of the partition
function in the form of a product of the partial partition functions
$\tilde Q_n$ of individual layers and the integral of the ``smoothed''
effective distribution of fluctuations. We have
\be
Z = 2^N 2^{(N_{n+1}-1)/2}
\tilde Q_0 \tilde Q_1 \cdots \tilde Q_n [Q(P_n)]^{N_{n+1}}
\int \tilde \cW_6^{(n+1)}(\rho) (\rd\rho)^{N_{n+1}}.
\label{cmp2d34}
\ee
The effective sextic distribution of fluctuations in the $(n+1)$th block
structure is written as follows:
\be
\tilde \cW_6^{(n+1)}(\rho) \! = \! \exp \!\! \lbr -\frac{1}{2}
\sum_{k\leqslant B_{n+1}} \tilde d_{n+1}(k) \rhok\rhomk -
\sum\limits_{l=2}^3 \frac{\tilde a_{2l}^{(n+1)}}
{(2l)!N_{n+1}^{l-1}} \sum_{k_1,\ldots,k_{2l}\leqslant B_{n+1}}
\rho_{\vk_1}\cdots\rho_{\vk_{2l}}
\delta_{\vk_1+\cdots+\vk_{2l}} \rbr \!\! .
\label{cmp2d34b}
\ee
Here, $B_{n+1}=B's^{-(n+1)}$, $N_{n+1}=N's^{-3(n+1)}$, and
$\tilde d_{n+1}(k)$, $\tilde a_4^{(n+1)}$, $\tilde a_6^{(n+1)}$ are
the renormalized values of coefficients $d'(k)$, $a'_4$, $a'_6$
after integration over $n+1$ layers of the phase space of CV.

In comparison with the results obtained earlier without taking into
account the correction for the potential averaging (see, for example,
\cite{ykpmo101,ykp202,p398,kpd297}), the expression (\ref{cmp2d34})
includes new quantities. In particular, the function
\be
f\left(h_n,\alpha_n\right) = - 48 s^{9/2} C^{1/2}(h_n,\alpha_n) F_2^3(\eta_n,\xi_n)
\frac{q_n t_n}{\sqrt{\tilde a_4^{(n)}}} \cF_0
\label{cmp2d37}
\ee
appearing in $\tilde Q_n$ characterizes the correction to the partial
partition functions. The basic arguments
$h_n=\tilde d_n\left(B_{n+1},B_n\right)\big(6/\tilde a_4^{(n)}\big)^{1/2}$ and
$\alpha_n=\sqrt 6 \tilde a_6^{(n)}\big/\big[15 \big(\tilde a_4^{(n)}\big)^{3/2}\big]$
are expressed in terms of the coefficients $\tilde d_n(B_{n+1},B_n)$
(the average value of $\tilde d_n(k)$ in the $n$th layer
of the CV phase space), $\tilde a_4^{(n)}$ and $\tilde a_6^{(n)}$.
The intermediate variables
$\eta_n=\left(6s^3\right)^{1/2}F_2(h_n,\alpha_n)\big/C^{1/2}(h_n,\alpha_n)$ and
$\xi_n=\sqrt 6 s^{-3/2}N(h_n,\alpha_n)\big/\big[15 C^{3/2}(h_n,\alpha_n)\big]$ are
functions of $h_n$ and $\alpha_n$. The special functions
$C(h_n,\alpha_n)=-F_4(h_n,\alpha_n)+3F_2^2(h_n,\alpha_n)$ and
$N(h_n,\alpha_n)=F_6(h_n,\alpha_n)-15F_4(h_n,\alpha_n)F_2(h_n,\alpha_n)+
30F_2^3(h_n,\alpha_n)$ are combinations of the
functions $F_{2l}(h_n,\alpha_n)=\linebreak I_{2l}(h_n,\alpha_n)\big/I_0(h_n,\alpha_n)$,
where $I_{2l}(h_n,\alpha_n)=\int_0^{\infty}t^{2l}
\exp\left(-h_n t^2-t^4-\alpha_n t^6\right) \rd t$.
The relation~(\ref{cmp2d37}) for \linebreak $f(h_n,\alpha_n)$, except
\be
q_n = q \frac{1+\alpha'_0}{s^2} \frac{1+\alpha'_1}{s^2}\cdots
\frac{1+\alpha'_{n-1}}{s^2}\, ,
\label{cmpdno4}
\ee
contains the factor
\be
t_n = \sqrt{\frac{\tilde a_4^{(n)}}{24}} \frac{1}{F_2(h_n,\alpha_n)}
\frac{s^2}{1+\alpha'_0} \frac{s^2}{1+\alpha'_1}\cdots
\frac{s^2}{1+\alpha'_{n-1}} t_0^{(n)} \frac{1}{\beta\tilde\Phi(0)}\, .
\label{cmp2d38}
\ee
For the quantity $\alpha'_n$, which defines the correction for
the averaging of the Fourier transform of the potential in the $n$th
layer of the phase space of CV, we obtain
\be
\alpha'_n = 144\pi^2 s^6 F_2^4(\eta_n,\xi_n) \bar q t_n \cB_0\, .
\label{cmp2d46}
\ee
The expressions for $\cF_0$ (appearing in $f(h_n,\alpha_n)$), $t_0^{(n)}$
(appearing in $t_n$) and $\cB_0$ (appearing in $\alpha'_n$) are
presented in~\cite{ykpmo101,ykp312}. At large values of the RG
parameter $s$, there emerge large intervals of wave vectors,
in which $\tilde{\Phi}(k)$ is averaged. In this case ($s>5$),
the correction $\beta\tilde{\Phi}(k)-\beta\tilde{\Phi}(B_{n+1},B_{n})$
is substantial, so that its accounting in the linear approximation
is incorrect.

\section{Analysis of recurrence relations for the
\texorpdfstring{$\rho^6$}{rho6} model. Critical exponent~\texorpdfstring{$\eta$}{eta}}

The coefficients $\tilde d_{n+1}(k)$, $\tilde a_4^{(n+1)}$ and
$\tilde a_6^{(n+1)}$ in (\ref{cmp2d34b}) satisfy the following RR:
\bea
 \tilde d_{n+1}(B_{n+2},B_{n+1}) &=&
(\tilde a_4^{(n)})^{1/2} \tilde Y(h_n,\alpha_n)
- q \frac{1+\alpha'_0}{s^2} \frac{1+\alpha'_1}{s^2}\cdots
\frac{1+\alpha'_{n-1}}{s^2} \lp 1 -
\frac{1+\alpha'_n}{s^2}\right), \nonumber\\
 \tilde a_4^{(n+1)} &=&
\tilde a_4^{(n)} s^{-3} \tilde B(h_n,\alpha_n), \non
\tilde a_6^{(n+1)} &=&
\left(\tilde a_4^{(n)}\right)^{3/2} s^{-6} \tilde D(h_n,\alpha_n).
\label{cmp2d40}
\eea
The contributions to the functions
\bea
 \tilde Y(h_n,\alpha_n)& =& Y(h_n,\alpha_n) \left[ 1 -
G(h_n,\alpha_n) \cA_0 \right], \nonumber\\
 \tilde B(h_n,\alpha_n) &=& B(h_n,\alpha_n) \left[ 1 +
\cK(h_n,\alpha_n) \cC_0 \right], \non
 \tilde D(h_n,\alpha_n) &=& D(h_n,\alpha_n) \left[ 1 -
\cL(h_n,\alpha_n) \cD_0 \right]
\label{cmp2d41}
\eea
from the potential averaging are given by the terms
$G(h_n,\alpha_n)\cA_0$, $\cK(h_n,\alpha_n)\cC_0$ and
$\cL(h_n,\alpha_n)\cD_0$. Here,
\bea
Y(h_n,\alpha_n) &=& s^{3/2} F_2(\eta_n,\xi_n)
C^{-1/2}(h_n,\alpha_n), \nonumber\\
B(h_n,\alpha_n) &=& s^6 C(\eta_n,\xi_n)
C^{-1}(h_n,\alpha_n), \non
D(h_n,\alpha_n) &=& s^{21/2} N(\eta_n,\xi_n)
C^{-3/2}(h_n,\alpha_n),
\label{cmp1d12}
\eea
and
\bea
G(h_n,\alpha_n)& =& 288 s^{9/2} F_2^3(\eta_n,\xi_n)
C^{1/2}(h_n,\alpha_n) \frac{q t_n}{\sqrt{\tilde u_n}}\, ,\nonumber \\
\cK(h_n,\alpha_n) &=& 1728 s^{3/2} \frac{F_2^5(\eta_n,\xi_n)}
{C(\eta_n,\xi_n)} C^{1/2}(h_n,\alpha_n)
\frac{q t_n}{\sqrt{\tilde u_n}}\, , \non
\cL(h_n,\alpha_n) &=& 17280 s^{-3/2} \frac{F_2^6(\eta_n,\xi_n)}
{N(\eta_n,\xi_n)} C^{1/2}(h_n,\alpha_n)
\frac{q t_n}{\sqrt{\tilde u_n}}\, .
\label{cmp2d42}
\eea
The quantities $\cA_0$, $\cC_0$ and $\cD_0$ as well as the quantities
$\cF_0$ and $\cB_0$ mentioned above appear due to the inclusion of
the averaging correction. They are calculated with the help of
the summation over the distances to the particles located at the lattice
sites (see~\cite{ykpmo101,ykp312}). In terms of the variables
\bea
\tilde r_n &=& \frac{s^2}{1+\alpha'_0}\frac{s^2}{1+\alpha'_1}\cdots
\frac{s^2}{1+\alpha'_{n-1}} \tilde d_n(0), \nonumber\\
\tilde u_n &=& \frac{s^4}{\left(1+\alpha'_0\right)^2}
\frac{s^4}{\left(1+\alpha'_1\right)^2}\cdots
\frac{s^4}{\left(1+\alpha'_{n-1}\right)^2} \tilde a_4^{(n)}, \non
\tilde w_n &=& \frac{s^6}{\left(1+\alpha'_0\right)^3}
\frac{s^6}{\left(1+\alpha'_1\right)^3}\cdots
\frac{s^6}{\left(1+\alpha'_{n-1}\right)^3} \tilde a_6^{(n)},
\label{cmp2d43}
\eea
the RR (\ref{cmp2d40}) assume the forms
\bea
\tilde r_{n+1} &=& \frac{s^2}{1+\alpha'_n} \lbr -q + \left(\tilde u_n\right)^{1/2}
\tilde Y(h_n,\alpha_n) \rbr, \nonumber\\
\tilde u_{n+1} &=& \frac{s}{\left(1+\alpha'_n\right)^2} \tilde u_n
\tilde B(h_n,\alpha_n), \non
\tilde w_{n+1} &=& \frac{1}{\left(1+\alpha'_n\right)^3} (\tilde u_n)^{3/2}
\tilde D(h_n,\alpha_n).
\label{cmp2d47}
\eea

There are two essential distinctions between the RR (\ref{cmp2d47}) and
those obtained without involving the correction for $\Delta\tilde\Phi(k)$~\cite{ykpmo101,p196,k189}. The first of them consists in the specific
substitution of variables~(\ref{cmp2d43}), which differs from
the corresponding substitution without the correction by including
the factors \linebreak $\left(1+\alpha'_0\right)\left(1+\alpha'_1\right)\cdots\left(1+\alpha'_{n-1}\right)$.
The second distinction concerns the transformation of special
functions $Y(h_n,\alpha_n)$, $B(h_n,\alpha_n)$ and $D(h_n,\alpha_n)$
(\ref{cmp1d12}) into the functions $\tilde Y(h_n,\alpha_n)$,
$\tilde B(h_n,\alpha_n)$ and $\tilde D(h_n,\alpha_n)$ (\ref{cmp2d41}).
This distinction is associated with a shift of the fixed-point
coordinates and with corrections to the critical exponents of
thermodynamic functions.

A particular solution of RR (\ref{cmp2d47}) is a new fixed point
($\tilde r, \tilde u, \tilde w$), which differs at
$\Delta\tilde\Phi(k)\neq 0$ from the fixed point
($r^{(0)}, u^{(0)}, w^{(0)}$) for the case of $\Delta\tilde\Phi(k)=0$
\cite{ykpmo101,p196,k189}. The coordinates of the fixed point of
RR (\ref{cmp2d47}) can be expressed as follows:
\be
\tilde r = - \tilde f \beta\tilde\Phi(0), \qquad
\tilde u = \tilde\varphi[\beta\Phi(0)]^2, \qquad
\tilde w = \tilde\psi[\beta\Phi(0)]^3.
\label{cmp3d7}
\ee
Here,
\bea
\tilde f &=& \bar q \left[\tilde Y\left(\tilde h,\tilde\alpha\right) - \tilde h/\sqrt 6\right]
\left[\tilde Y\left(\tilde h,\tilde\alpha\right) - \left(1 + \alpha^{\prime(0)}\right)
\tilde h\Big/\left(s^2 \sqrt 6\right)\right]^{-1}, \nonumber\\
\tilde\varphi & =& \bar q^2 \left[1 - \left(1 + \alpha^{\prime(0)}\right) s^{-2}\right]^2
\left[\tilde Y\left(\tilde h,\tilde\alpha\right) - \left(1 + \alpha^{\prime(0)}\right)
\tilde h\Big/\left(s^2 \sqrt 6\right)\right]^{-2}, \non
\tilde\psi &=& \left(1 + \alpha^{\prime(0)}\right)^{-3} \left(\tilde\varphi\right)^{3/2}
\tilde D\left(\tilde h,\tilde\alpha\right),
\label{cmp3d8}
\eea
and
\bea
 \tilde h &=& \sqrt 6 \frac{\tilde r + q}{(\tilde u)^{1/2}}\,, \nonumber\\
 \tilde\alpha &=&\frac{\sqrt 6}{15} \frac{\tilde w}{(\tilde u)^{3/2}}\,, \non
 \alpha^{\prime(0)} &=& 144\pi^2 s^6 F_2^4\left(\eta^{(0)},\xi^{(0)}\right)
\bar q t^{(0)} \cB_0\,, \non
 t^{(0)} &=& t_n (u^{(0)}, h^{(0)}, \alpha^{(0)})\,.
\label{cmp3d4p6p9}
\eea
The quantities $h^{(0)}$, $\alpha^{(0)}$ and $\eta^{(0)}$, $\xi^{(0)}$
describe, respectively, the basic ($h_n$, $\alpha_n$) and
intermediate ($\eta_n$, $\xi_n$) arguments at the fixed
point obtained without involving the correction for the potential
averaging. In the linear approximation in $\Delta\tilde\Phi(k)$,
we have
\bea
\tilde f& = &f_0 \Bigg( 1 + \Big\{\left[Y'_h\left(h^{(0)},\alpha^{(0)}\right)
h^{(0)}\big/\sqrt 6 - Y\left(h^{(0)},\alpha^{(0)}\right)\big/\sqrt 6\right] \Delta h +
Y'_\alpha\left(h^{(0)},\alpha^{(0)}\right) h^{(0)}\big/\sqrt 6 \Delta \alpha \nonumber\\
& & - Y\left(h^{(0)},\alpha^{(0)}\right) G\left(h^{(0)},\alpha^{(0)}\right)
\cA_0 h^{(0)}\big/\sqrt 6\Big\}
\left(1 - s^{-2}\right) \Big\{\left[Y\left(h^{(0)},\alpha^{(0)}\right) - h^{(0)}\big/\sqrt 6\right] \non
& & \times \left[Y\left(h^{(0)},\alpha^{(0)}\right) - h^{(0)}\big/\big(s^2 \sqrt 6\big)\right]\Big\}^{-1} +\alpha^{\prime(0)} h^{(0)}\big/\big(s^2 \sqrt 6\big)
\left[Y\left(h^{(0)},\alpha^{(0)}\right) - h^{(0)}\big/\big(s^2 \sqrt 6\big)\right]^{-1} \Bigg), \non 
\tilde\varphi &=& \varphi_0 \Bigg( 1 \! + \!
2\Big\{-\left[Y'_h\left(h^{(0)},\alpha^{(0)}\right) \! - \! 1\big/\big(s^2 \sqrt 6\big)\right] \Delta h \! - \! Y'_\alpha\left(h^{(0)},\alpha^{(0)}\right) \Delta \alpha \! + \!
Y\left(h^{(0)},\alpha^{(0)}\right) G(h^{(0)},\alpha^{(0)}) \cA_0 \non
& & + \alpha^{\prime(0)} h^{(0)}\big/\big(s^2 \sqrt 6\big)\Big\}
\left[Y\left(h^{(0)},\alpha^{(0)}\right) - h^{(0)}\big/\big(s^2 \sqrt 6\big)\right]^{-1} -
2\alpha^{\prime(0)} s^{-2}\big/\big(1 - s^{-2}\big) \Bigg), \non
\tilde\psi& = &\psi_0 \Bigg[1 +
3\Bigg(\Big\{\left[Y\left(h^{(0)},\alpha^{(0)}\right) - h^{(0)}\big/\big(s^2 \sqrt 6\big)\right]
D'_h\left(h^{(0)},\alpha^{(0)}\right)\big/3 -
\left[Y'_h\left(h^{(0)},\alpha^{(0)}\right) - 1\big/\big(s^2 \sqrt 6\big)\right] \non
& & \times D\left(h^{(0)},\alpha^{(0)}\right)\Big\} \Delta h +
\Big\{\left[Y\left(h^{(0)},\alpha^{(0)}\right) - h^{(0)}\big/\big(s^2 \sqrt 6\big)\right]
D'_\alpha\left(h^{(0)},\alpha^{(0)}\right)\big/3 \nonumber \\
& & - Y'_\alpha\left(h^{(0)},\alpha^{(0)}\right) D\left(h^{(0)},\alpha^{(0)}\right)\Big\}
\Delta \alpha + Y\left(h^{(0)},\alpha^{(0)}\right) D\left(h^{(0)},\alpha^{(0)}\right)
G\left(h^{(0)},\alpha^{(0)}\right) \cA_0 \non
& & + \alpha^{\prime(0)} h^{(0)} D(h^{(0)},\alpha^{(0)})\big/\big(s^2 \sqrt 6\big)\Bigg)
\Big\{\left[Y\left(h^{(0)},\alpha^{(0)}\right) - h^{(0)}\big/\big(s^2 \sqrt 6\big)\right]
D\left(h^{(0)},\alpha^{(0)}\right)\Big\}^{-1} \non
& & - 3\alpha^{\prime(0)}\big/\left(1 - s^{-2}\right) -
\cL\left(h^{(0)},\alpha^{(0)}\right) \cD_0 \Bigg].
\label{cmp3d10}
\eea
The quantities $f_0$, $\varphi_0$ and $\psi_0$ characterize the
fixed-point coordinates in the case when $\eta=0$. Formulas for the
derivatives $Y'_h\left(h^{(0)},\alpha^{(0)}\right)$,
$Y'_\alpha\left(h^{(0)},\alpha^{(0)}\right)$, $D'_h\left(h^{(0)},\alpha^{(0)}\right)$ and
$D'_\alpha(h^{(0)},\alpha^{(0)})$ can be written using series expansions
of the corresponding functions in the vicinity of the fixed
point~\cite{p196}. The differences $\Delta h=\tilde h-h^{(0)}$
and $\Delta \alpha=\tilde \alpha-\alpha^{(0)}$ determine the
displacements of the basic arguments $h_n$ and $\alpha_n$ at the fixed
points ($\tilde r, \tilde u, \tilde w$)
and $\left(r^{(0)}, u^{(0)}, w^{(0)}\right)$.

The RR (\ref{cmp2d47}) make it possible to find the elements of the RG
linear transformation matrix. These matrix elements $\tilde R_{ij}$
($i=1,2,3$; $j=1,2,3$) can be presented in the following
forms (the linear approximation in $\Delta\tilde\Phi(k)$):
\bea
 \tilde R_{11} &=& R_{11} \left(1-\alpha^{\prime(0)}\right) +
R_{11}^{(1h)} \Delta h + R_{11}^{(1\alpha)} \Delta \alpha +
R_{11}^{(2)} \cA_0\,, \nonumber\\
 \tilde R_{22} &=& R_{22} \left(1-2\alpha^{\prime(0)}\right) +
R_{22}^{(1h)} \Delta h + R_{22}^{(1\alpha)} \Delta \alpha +
R_{22}^{(2)} \cC_0\,, \non
 \tilde R_{33} &=& R_{33} \left(1-3\alpha^{\prime(0)}\right) +
R_{33}^{(1h)} \Delta h + R_{33}^{(1\alpha)} \Delta \alpha +
R_{33}^{(2)} \cD_0\,, \non
 \tilde R_{ij} &=& \tilde R_{ij}^{(0)} \left(\tilde u\right)^{(i-j)/2}\,, \qquad
i\neq j\,, \non
 \tilde R_{1k_1}^{(0)} &=& R_{1k_1}^{(0)} \left(1-\alpha^{\prime(0)}\right) +
R_{1k_1}^{(1h)} \Delta h + R_{1k_1}^{(1\alpha)} \Delta \alpha +
R_{1k_1}^{(2)} \cA_0\,, \qquad \ \;k_1 = 2,3\,, \non
 \tilde R_{2k_2}^{(0)} &=& R_{2k_2}^{(0)} \left(1-2\alpha^{\prime(0)}\right) +
R_{2k_2}^{(1h)} \Delta h + R_{2k_2}^{(1\alpha)} \Delta \alpha +
R_{2k_2}^{(2)} \cC_0\,, \qquad k_2 = 1,3\,, \non
 \tilde R_{3k_3}^{(0)} &=& R_{3k_3}^{(0)} \left(1-3\alpha^{\prime(0)}\right) +
R_{3k_3}^{(1h)} \Delta h + R_{3k_3}^{(1\alpha)} \Delta \alpha +
R_{3k_3}^{(2)} \cD_0\,, \qquad k_3 = 1,2\,.
\label{cmp3d15p18p20p23}
\eea
It should be noted that formulas for the quantities $R_{ii}$ and
$R_{ij}^{(0)}$ ($i\neq j$) from (\ref{cmp3d15p18p20p23}) coincide with
the corresponding expressions for the matrix elements obtained
without taking into account the correction for the potential
averaging~\cite{ykpmo101}. The contributions to the matrix elements
$\tilde R_{ij}$ from terms $R_{ij}^{(1h)}\Delta h$ and
$R_{ij}^{(1\alpha)}\Delta \alpha$ correspond to a fixed-point shift
due to the inclusion of the dependence of the Fourier transform of
the interaction potential on the wave vector. The terms like
$R_{ij}^{(2)}\cA_0$, $R_{ij}^{(2)}\cC_0$ and $R_{ij}^{(2)}\cD_0$
describe a direct contribution to $\tilde R_{ij}$ from
the correction for averaging.

The explicit solutions of RR
\bea
 \tilde r_n &=& \tilde r + \tilde c_1 \tilde E_1^n +
\tilde c_2 \tilde w_{12}^{(0)} (\tilde u)^{-1/2} \tilde E_2^n +
\tilde c_3 \tilde w_{13}^{(0)} (\tilde u)^{-1} \tilde E_3^n\,,\nonumber \\
 \tilde u_n &=& \tilde u +
\tilde c_1 \tilde w_{21}^{(0)} (\tilde u)^{1/2} \tilde E_1^n +
\tilde c_2 \tilde E_2^n +
\tilde c_3 \tilde w_{23}^{(0)} (\tilde u)^{-1/2} \tilde E_3^n\,, \non
 \tilde w_n &=& \tilde w +
\tilde c_1 \tilde w_{31}^{(0)} \tilde u \tilde E_1^n +
\tilde c_2 \tilde w_{32}^{(0)} (\tilde u)^{1/2} \tilde E_2^n +
\tilde c_3 \tilde E_3^n
\label{cmp3d25}
\eea
in terms of (\ref{cmp2d43}) read
\bea
 \tilde d_n(B_{n+1},B_n) &=& s^{-2n} \lbr \prod_{m=0}^{n-1}
\left(1+\alpha'_m\right) \rbr \lbr \tilde r + q + \tilde c_1 \tilde E_1^n +
\tilde c_2 \tilde w_{12}^{(0)} (\tilde u)^{-1/2} \tilde E_2^n +
\tilde c_3 \tilde w_{13}^{(0)} (\tilde u)^{-1} \tilde E_3^n \rbr, \nonumber\\
 \tilde a_4^{(n)} &=& s^{-4n} \lbr \prod_{m=0}^{n-1}
\left(1+\alpha'_m\right)^2 \rbr \lbr \tilde u +
\tilde c_1 \tilde w_{21}^{(0)} (\tilde u)^{1/2} \tilde E_1^n +
\tilde c_2 \tilde E_2^n +
\tilde c_3 \tilde w_{23}^{(0)} (\tilde u)^{-1/2} \tilde E_3^n \rbr, \non
 \tilde a_6^{(n)} &=& s^{-6n} \lbr \prod_{m=0}^{n-1}
\left(1+\alpha'_m\right)^3 \rbr \lbr \tilde w +
\tilde c_1 \tilde w_{31}^{(0)} \tilde u \tilde E_1^n +
\tilde c_2 \tilde w_{32}^{(0)} (\tilde u)^{1/2} \tilde E_2^n +
\tilde c_3 \tilde E_3^n \rbr.
\label{cmp3d29}
\eea
Here, $\tilde r$, $\tilde u$ and $\tilde w$ are given in (\ref{cmp3d7})
and (\ref{cmp3d10}). The coefficients $\tilde c_l$ are obtained from the
initial conditions at $n=0$. The temperature-independent quantities
$\tilde w_{il}^{(0)}$ determine the eigenvectors of the RG linear
transformation matrix, and $\tilde E_l$ are the eigenvalues
of this matrix.

The main distinction between the solutions (\ref{cmp3d29}) and
the solutions of RR in the absence of the correction for the potential
averaging~\cite{ykpmo101,ykp202} lies in the availability of factors
of the type $\left(1+\alpha'_m\right)$. Taking into account the fact that
$\lim_{m\rightarrow\infty} \alpha'_m (T_\mathrm{c})=\alpha^{\prime(0)}$ at
$T=T_\mathrm{c}$, we obtain the following asymptotics in $n$ for the quantities
$\tilde d_n$, $\tilde a_4^{(n)}$ and $\tilde a_6^{(n)}$
from (\ref{cmp3d29}):
\bea
 \tilde d_n(B_{n+1},B_n) &=& (\tilde r + q) s^{-n(2-\eta)}, \nonumber\\
 \tilde a_4^{(n)} &=& \tilde u s^{-2n(2-\eta)}, \non
 \tilde a_6^{(n)} &=& \tilde w s^{-3n(2-\eta)}.
\label{cmp3d30}
\eea
The quantity $\eta$ is given by the formula
\be
\eta = \frac{\alpha^{\prime(0)}}{\ln s}
\label{cmp3d31}
\ee
and corresponds to the critical exponent of the correlation function.
Thus, the correction for the potential
averaging being involved in calculating the partition function of the system,
leads to a change of the asymptotics for the coefficients $\tilde d_n$,
$\tilde a_4^{(n)}$ and $\tilde a_6^{(n)}$ at $T=T_\mathrm{c}$ (in contrast to
the case of $\Delta\tilde\Phi(k)=0$, the exponents of these coefficients
contain the quantity $\eta$).

The renormalization of the critical exponent of the correlation length
$\nu=\ln s/\ln\tilde E_1$, compared to the case of $\eta=0$, is
associated with a change of the larger eigenvalue ($\tilde E_1>1$) of
the RG linear transformation matrix. In contrast to $\nu$, the critical
exponent of the susceptibility $\gamma=(2-\eta)\nu$ explicitly depends
on $\eta$. The specific heat of the system is characterized by
the exponent $\alpha=2-3\nu$, the expression for which contains a
renormalized critical exponent of the correlation length $\nu$.

\section{Discussion and conclusions}

Within the CV approach, an analytic method for calculating the free
energy of a 3D Ising-like system near the critical point was
elaborated in~\cite{ykp312} with the allowance for the simplest non-Gaussian
fluctuation distribution (the $\rho^4$ model) and the critical exponent
$\eta$. The critical exponent of the correlation function $\eta=0.024$
was found in~\cite{ykp312} by including the correction for the potential
averaging in the course of calculating the partition function of
the system. This value of $\eta$ is in accord with the other authors' data.
For comparison, the exponents $\eta=0.0335(25)$, $\eta=0.0362(8)$
and $\eta=0.033$ were obtained within the framework of the field-theory
approach (7-loop calculations)~\cite{gz298}, Monte Carlo
simulations~\cite{h101} and non-perturbative RG approach (the order
$\partial^4$ of the derivative expansion)~\cite{cdm203}, respectively.
Some difference between the value of $\eta=0.024$ and other authors'
data can be connected with the approximations in calculations,
in particular, with the approximations which are made within
the CV method (the simplest non-Gaussian distribution is used
for obtaining $\eta=0.024$; the correction inserted by the operator
$\hat\Delta_g$ is considered in the linear approximation in
$\Delta\tilde\Phi(k)$; the terms proportional to the higher orders of
operators $\partial/\partial \rhok$ are not taken into account in
the expression for $\hat\Delta_g$; the operator $\hat\Delta_g$ in
(\ref{cmp2d3}) acts only on the first term in the exponent).
As is established in~\cite{ykp312}, the inclusion of a nonzero exponent
$\eta$ within the CV method leads to a reduction of the value of
the critical exponent for the correlation length, $\nu$ (in comparison
with the case of $\eta=0$). For better quantitative estimates
of $\nu$ and other renormalized critical exponents, it is necessary
to use the non-Gaussian approximations higher than the quartic
distribution (e.g., the sextic one).

This paper supplements the previous study~\cite{ykp312} based on
the $\rho^4$ model. In the present publication, the correction for
the averaging of the Fourier transform of the interaction potential
is taken into account using the sextic fluctuation
distribution (the $\rho^6$ model). The effect of the mentioned
correction on the critical behaviour of a 3D uniaxial magnet is
investigated in the linear approximation.

Extending the method for the layer-by-layer integration of the partition
function of the system to the case of the $\rho^6$ model, the RR for
the coefficients of the effective distributions are written and
analysed. It is shown that the inclusion of the correction for the potential
averaging gives rise to a change of the asymptotics for the RR solutions
at $T=T_\mathrm{c}$.

The critical exponents of the correlation length, susceptibility and
specific heat are renormalized due to the above-mentioned correction.

Our analytic representations acquire a more general and complete
character as compared with the case when $\eta=0$.
An explicit expression (\ref{cmp2d34}) for the partition function
allows us to calculate the free energy and other thermodynamic
characteristics of the system near $T_\mathrm{c}$ taking into account
the non-Gaussian sextic distribution and the small critical
exponent $\eta$.

An analytic procedure for the calculation of the critical exponent
of the correlation function developed in this paper on the basis of
the $\rho^6$ model for a one-component spin system may be generalized
to the case of a system with an $n$-component order parameter.



\newpage

\ukrainianpart

\title{Критична поведінка тривимірної ізингоподібної системи \\
в наближенні моделі $\rho^6$: Роль поправки  на усереднення потенціалу}
\author{І.В. Пилюк\refaddr{label1}, М.В. Уляк\refaddr{label2}}
\addresses{
\addr{label1} Інститут фізики конденсованих систем НАН України,
вул. Свєнціцького, 1, 79011 Львів, Україна
\addr{label2} Державний вищий навчальний заклад ``Гірничо-економічний
коледж'',  \\ вул. Стуса, 17, 80100 Червоноград, Україна
}
%
%
%

\makeukrtitle

\begin{abstract}
\tolerance=3000%
З використанням методу колективних змінних (КЗ) вивчається критична
поведінка систем, які належать до класу універсальності тривимірної
моделі Ізинга. Статистична сума однокомпонентної спінової системи
обчислюється шляхом інтегрування за шарами фазового простору КЗ в
наближенні негаусового шестирного розподілу флуктуацій параметра
порядку (модель $\rho^6$). Особливістю запропонованого розрахунку є
прийняття до уваги залежності фур'є-образу потенціалу взаємодії від
хвильового вектора. Врахування поправки на усереднення потенціалу
приводить до відмінного від нуля критичного показника кореляційної
функції $\eta$ і перенормування значень інших критичних показників.
Виділено внески від цієї поправки в рекурентні співвідношення для
моделі $\rho^6$, координати фіксованої точки та елементи матриці
лінійного перетворення ренормалізаційної групи. Вираз для малого
критичного показника $\eta$ отримано у вищому негаусовому наближенні.
\keywords тривимірна ізингоподібна система, критична поведінка,
шестирний розподіл, усереднення потенціалу, малий критичний показник

\end{abstract}

\end{document}